\documentclass[twocolumn,prl,showpacs,floatfix]{revtex4}
\usepackage{graphicx}
\usepackage{epsfig}
\usepackage{rotating}
\usepackage{amsmath}
\usepackage{amsfonts}
\usepackage{amssymb}
\usepackage{enumerate}
\usepackage{longtable}
\usepackage{dcolumn}% Align table columns on decimal point
\usepackage{bm}

\begin{document}

\title{Incipient and well-developed entropy plateaus in spin-S Kitaev models
}

\author{J. Oitmaa }
\affiliation{School of Physics, The University of New South Wales,
Sydney 2052, Australia}

\author{A. Koga}
\affiliation{Department of Physics, Tokyo Institute of Technology, Meguro, Tokyo, 152-8551, Japan}

\author{R. R. P. Singh}
\affiliation{Department of Physics, University of California Davis, CA 95616, USA}

\date{\rm\today}

\begin{abstract}
We present results on entropy and heat-capacity of the spin-S honeycomb-lattice Kitaev models
using high-temperature series expansions and thermal pure quantum (TPQ) state methods. We study
models with anisotropic couplings $J_z=1\ge J_x=J_y$ for spin values 1/2, 1, 3/2, and 2. We show
that for $S>1/2$, any anisotropy leads to well developed plateaus in the entropy function at an entropy value
of $\frac{1}{2}\ln{2}$, independent of $S$. However, in the absence of anisotropy, there is an incipient entropy
plateau at $S_{max}/2$, where $S_{max}$ is the infinite temperature entropy of the system. We discuss possible
underlying microscopic reasons for the origin and implications of these entropy plateaus.

\end{abstract}


\maketitle

\section{Introduction}
In frustrated magnets the existence of residual entropy at temperatures well below the development
of short-range order and multiple peaks in the heat capacity as a function of temperature are well known 
phenomena \cite{balents}.
The theoretical basis for these date back to the works of Pauling \cite{pauling} on residual entropy of ice and and Wannier's
exact solution of the triangular-lattice Ising antiferromagnet \cite{wannier}. Experimentally, 
entropy plateaus are best known in the spin-ice materials \cite{spin-ice}. Such a behavior reflects the existence of a low energy manifold in the system, whose size and nature is intimately linked to the spin-liquid phase.

In a pioneering but relatively unheralded paper Baskaran, Sen and Shankar (BSS) \cite{bss} 
considered the spin-S generalization
of the celebrated Kitaev's spin-half Honeycomb model \cite{kitaev}. They showed that even though the models with spin greater than
half are no longer exactly soluble, they retain many of the features of the spin-half model. Regardless of spin, one 
can define loop operators on elementary hexagons that commute with each other and with the Hamiltonian, thus defining
an infinite number of conserved $Z_2$-valued fluxes. In the classical limit, there is an exponentially large number of
ground states. 
%characterized by continuous parameters defined on closed loops. 
Of these, the so called Cartesian States 
represent a finite entropy manifold which is favored by quantum fluctuations. Later work by Chandra, Ramola and
Dhar \cite{crd} and by Rousochatzakis, Sizyuk and Perkins \cite{rsp} has further explored the ground-state manifold of the
classical and large-S limits of the model
and even found a mapping back to the spin-half Kitaev model through a sequence of intricate selections in the
degenerate subspaces.

\begin{figure}
\begin{center}
 \includegraphics[width=8cm]{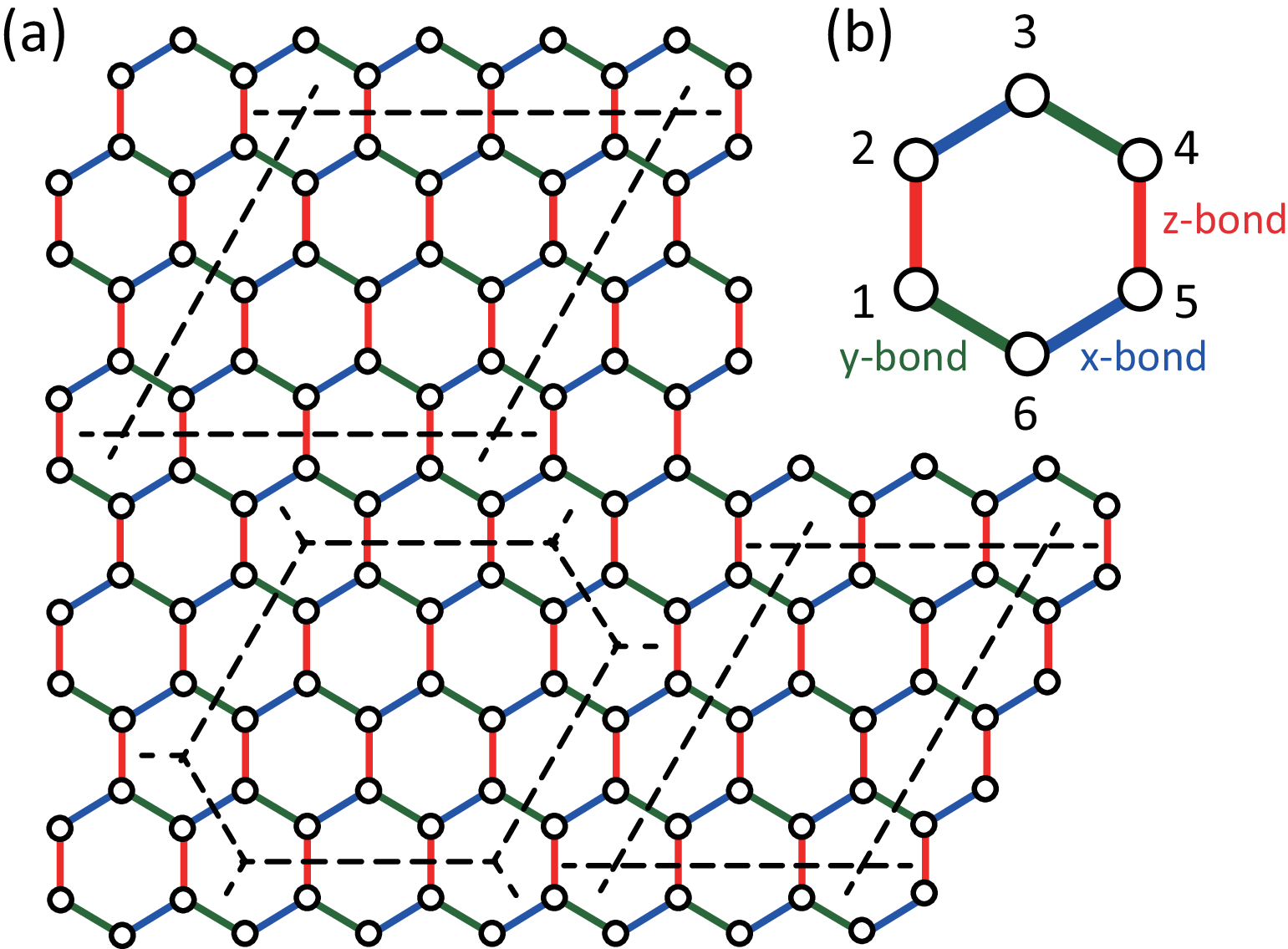}
\caption{\label{fig1} 
1a. Honeycomb lattice with $J_x$, $J_y$ and $J_z$ bonds denoted by different colors. 1b. An elementary plaquette 
with spins labelled 1 through 6.
}
\end{center}
\end{figure}

In a more recent study one of the authors of this work together with Tomishige and Nasu \cite{ktn} explored 
numerically the ground state 
and thermodynamic properties of the Kitaev model with varying-$S$ on finite systems using thermal pure quantum \cite{tpq1,tpq2,tpq}
and Monte Carlo \cite{qmc1,qmc2} methods,
where evidence was presented for an incipient entropy plateau
at a value of $\frac{1}{2}S_{max}$ for $S\le 2$, where $S_{max}$ is the infinite temperature entropy of the system. The purpose of this work is to follow up that study with high temperature expansions (HTE) \cite{book,so}
as well as thermal pure quantum (TPQ) state \cite{tpq1,tpq2,tpq} calculations. 
We study various S values as well as allow for an
anisotropy in the Kitaev couplings $J_z=1\ge J_x=J_y$. We confirm that there is an incipient entropy plateau
in the model in the absence of anisotropy at an entropy value of $S_{max}/2$. However, any anisotropy in the larger S systems drives them to
a well defined entropy plateau which occurs at a value of $\frac{1}{2}\ln{2}$. For $S=1/2$, $S_{max}/2$ and $\frac{1}{2}\ln{2}$ are the same
but they become further and further apart as the value of the spin increases.

The degeneracies of the anisotropic models are  easily understood in terms of the ground state degeneracy
of the classical model which increases as $2^{N/2}$ for an N-site system. One would expect this result to remain
valid, at least for large-S, because of the gap to remaining states, which scales as $JS$.
In contrast, for the isotropic case
our results imply a low energy manifold of $(2S+1)^{N/2}$ states with varying $S$. We present arguments that such a low energy
manifold, in the large-S limit, may arise from the continuous degeneracies present in the classical limit.

\section{Models and Methods}
We study the spin-$S$ honeycomb-lattice Kitaev model with Hamiltonian
\begin{equation}
{\cal H}=J_z\sum_{<i,j>} S_i^z S_j^z
+J_x\sum_{(i,k)} S_i^x S_k^x
+J_y\sum_{[i,l]} S_i^y S_l^y,
\end{equation}
 where the nearest-neighbors $<i,j>$, $(i,k)$ and $[i,l]$ point along the three different bond directions
of the honeycomb lattice respectively. The spin-operators correspond to a spin-value of $S$. 
We set $J_z=1$
and take $J_x=J_y\le 1$. 
%Fig.~1a shows a part of the honeycomb lattice, with finite periodic clusters used in
%the thermal pure quantum (TPQ) method based calculations enclosed by dashed lines. Fig.~1b shows an elementary plaquette.

\begin{figure}
\begin{center}
 \includegraphics[width=6cm]{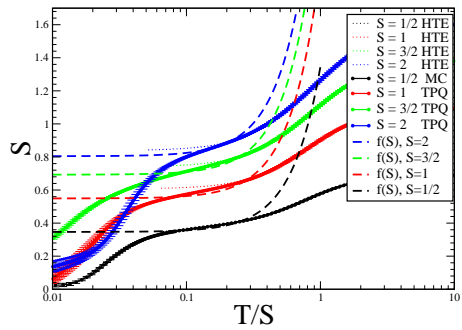}
\caption{\label{fig2} 
Entropy of the isotropic spin-S Kitaev model for various S values.
The limiting value of $S_{max}/2$ is indicated by the fits to the function
$f(S)=\ln{(2S+1)}/ 2 +\  \sqrt{S} \  (T/S)^2$ for different $S$ values. The spin-half Monte Carlo simulation data is provided by Nasu {\it et al} \cite{qmc1,qmc2}.
}
\end{center}
\end{figure}

\begin{figure}
\begin{center}
 \includegraphics[width=6cm]{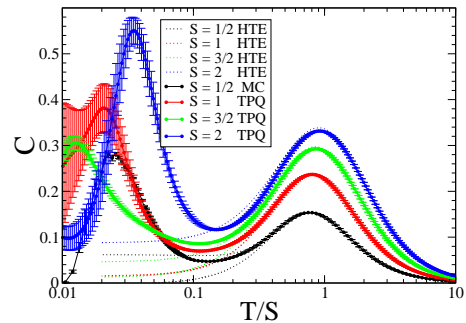}
\caption{\label{fig3} 
Specific heat of the isotropic spin-S Kitaev model for various S values.
The spin-half Monte Carlo simulation data is provided by Nasu {\it et al} \cite{qmc1,qmc2}.
}
\end{center}
\end{figure}

For the Kitaev models, the energy spectrum is identical under the change of sign of all $J_s$.
Consequently, the high temperature series expansions are even functions of $\beta$. High temperature
expansion coefficients are unique for the model and can be computed by several different methods.
Here we use the linked cluster methods to calculate them \cite{book}. Series expansions are
computed for the logarithm of the partition function from which the expansions for entropy, internal energy
and heat capacity follow. The expansions are carried out to order $\beta^{16}$ for $S=1/2$ and $S=1$, to
order $\beta^{14}$ for $S=3/2$ and $S=2$ and to order $\beta^{12}$ for $S=5/2$. The entropy series are
extrapolated using Pad\'e and differential approximants \cite{book}. The convergence is poorer for the heat capacity
series than for the entropy, as is usually the case, with the latter being a temperature derivative of the former.

We also use thermal pure quantum (TPQ) states \cite{tpq1,tpq2}
for calculating thermodynamic properties in the system.
A TPQ state at $T\rightarrow\infty$ is simply given by a random vector,
\begin{eqnarray}
|\psi_0\rangle=\sum_ic_i|i\rangle
\end{eqnarray}
where $|i\rangle(=|m_1\rangle\otimes |m_2\rangle \otimes\cdots \otimes|m_N\rangle)$ is represented by a direct product of
the local eigenstates $|m_i\rangle$ of $S_i^z$
with eigenvalue $m(=-S, -S+1, \cdots, S)$ at site $i$ and ${c_i}$ is a set
of random complex numbers under the normalized contraint.
By multiplying the Hamiltonian by a certain TPQ state,
the TPQ states at lower temperatures are constructed.
The $k$th TPQ state is represented as
\begin{eqnarray}
|\psi_k\rangle&=&\frac{(l-{\cal H})|\psi_{k-1}\rangle}{|(l-{\cal H})|\psi_{k-1}\rangle|},
\end{eqnarray}
where $l$ is a constant value, which is larger than
the maximum eigenvalue of the Hamiltonian.
The corresponding inverse temperature is given by
\begin{eqnarray}
\beta_k&=&\frac{2k}{l-\langle \psi_k|{\cal H}|\psi_k\rangle}.
\end{eqnarray}
The specific heat $C$ and entropy $S$ are given by the following formula as
\begin{eqnarray}
C&=&\frac{dE}{dT},\\
S&=&\ln (2S+1) - \int_T^\infty \frac{C}{T'}dT',
\end{eqnarray}
where $E$ is the internal energy per site.

This method is formaly exact in the thermodynamic limit $N\rightarrow \infty$.
%and even in the finite system, the method has successfully been applied 
%to the quantum spin systems such as ...
When the TPQ state method is applied to the finite size system,
thermodynamic quantities depend on the system size and initial states.
As for the higher temperature peak in the Kitaev model,
the characteristic energy scale is large and thereby
the relatively smaller systems can capture thermodynamic properties correctly \cite{ktn,yamaji}.
In addition, the sample dependence of thermodynamic quantities
obtained by several TPQ states
is not so large in the temperatures
(its statistical errors are explicitly shown in the figures).
These allow us to apply the TPQ state method to the generalized Kitaev model.

\section{Numerical Results for entropy and heat capacity}

{We present numerical
results for $S\le 2$ comparing results of HTE with those obtained with the TPQ method \cite{tpq1,tpq2}. 
}
To examine the thermodynamic quantities at finite temperatures by the TPQ state approach,
we treat clusters with $N=$ $18$, $16$, and $12$ for spin $1$, $3/2$ and $2$ respectively [see Fig.~1].
We prepare, in each case, more than $10$ random vectors for the initial states, and the physical quantities
are calculated by averaging the values generated by these initial states. The method also allows calculation
of uncertainties in the physical quantities \cite{tpq1,tpq2}.
We have also done HTE for S=5/2 and do not see any qualitative change in going to this higher spin value.

In Fig.~2, we show the results for the entropy of the isotropic model $J_z=J_x=J_y$ for different $S$. 
The results for HTE and TPQ are shown. 
The temperature axis is scaled by $JS$. 
We have also plotted the function $f(S) = \ln{(2S+1)}/2 + \sqrt{S} \ (T/S)^2$ for comparison. One
can see the flattening of the entropy curves around $S_{max}/2 = \ln{(2S+1)}/2$. 
In this scaled temperature variable the incipient plateau arises at comparable values for different $S$.
The agreement between HTE and TPQ confirms that the results are accurate in the thermodynamic limit.

\begin{figure}
\begin{center}
 \includegraphics[width=6cm]{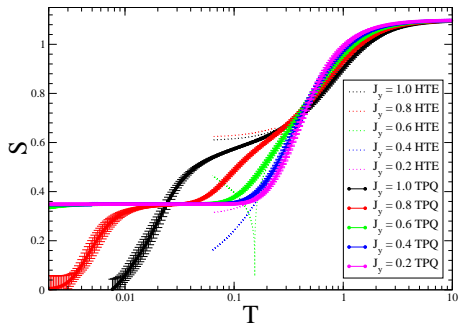}
\caption{\label{fig4} 
Entropy of the anisotropic spin-1 Kitaev model
}
\end{center}
\end{figure}

\begin{figure}
\begin{center}
 \includegraphics[width=6cm]{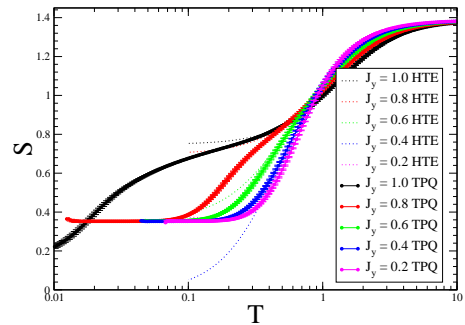}
\caption{\label{fig5} 
Entropy of the anisotropic spin-3/2 Kitaev model
}
\end{center}
\end{figure}

\begin{figure}
\begin{center}
 \includegraphics[width=6cm]{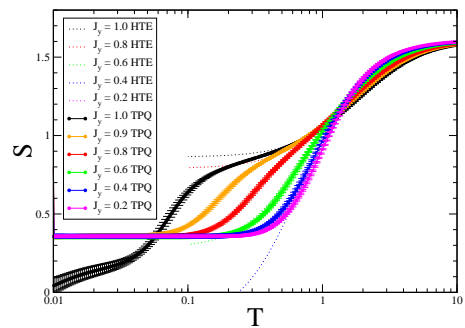}
\caption{\label{fig6} 
Entropy of the anisotropic spin-2 Kitaev model
}
\end{center}
\end{figure}

In Fig.~3, the specific heat of the isotropic model for different $S$ values is shown with the temperature
axis scaled by $JS$. It is clear that there is excellent agreement between HTE and TPQ data at high temperatures.
The HTE convergence starts to break down below the high temperature peak. But, TPQ results are valid down to lower temperatures.
A multi-peaked specific heat as a function of temperature is evident just from the fact that a significant amount
of entropy still has to be removed from the system below the high temperature peak \cite{suzuki}.

Fig.~4 through Fig.~6 show the entropy of the anisotropic models for $S=1$, $S=3/2$ and $S=2$.
It is clear that in the anisotropic models there is a well defined entropy plateau precisely at an entropy value of
$\frac{1}{2}\ln{2}$ regardless of the spin value. We should note that we do not expect strict plateaus in the entropy at finite temperatures as that would make it a non-analytic function of temperature. But, from a numerical point of view, the behavior seems indistinguishable from a plateau. Only for weak and zero anisotropy there is an incipient plateau in the entropy
near $S_{max}/2$. The entropy plateaus are very well developed in the anisotropic models as seen from the figures.
As $S$ increases the flattening near $S_{max}/2$ occurs closer and closer to $J_y=1$. The results are consistent with
the idea that in the large-S limit,
any anisotropy eliminates the incipient plateau near $S_{max}/2$ and only leaves an entropy plateau at $\frac{1}{2}\ln{2}$. 
It is also clear that at zero anisotropy ($J_x=J_y=J_z$), there is no plateau in the entropy at $\frac{1}{2}\ln{2}$ for any spin greater than
one half.

\begin{figure}
\begin{center}
 \includegraphics[width=6cm]{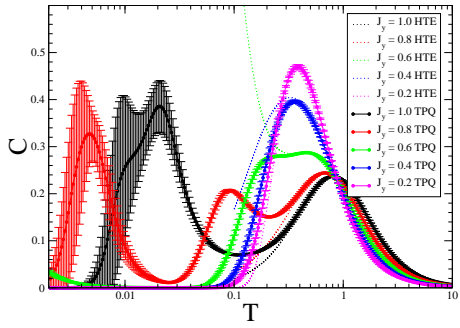}
\caption{\label{fig7} 
Specific heat of the anisotropic spin-1 Kitaev model
}
\end{center}
\end{figure}

\begin{figure}
\begin{center}
 \includegraphics[width=6cm]{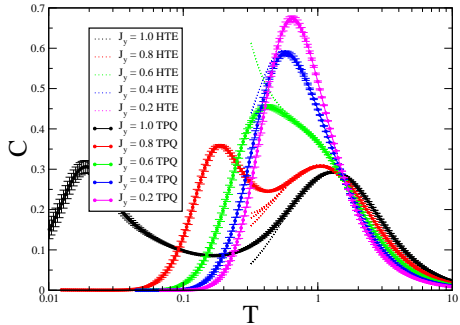}
\caption{\label{fig8} 
Specific heat of the anisotropic spin-3/2 Kitaev model
}
\end{center}
\end{figure}

\begin{figure}
\begin{center}
 \includegraphics[width=6cm]{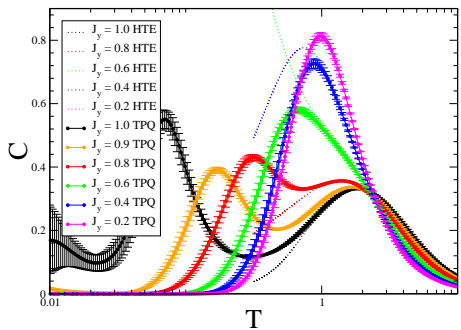}
\caption{\label{fig9} 
Specific heat of the anisotropic spin-2 Kitaev model
}
\end{center}
\end{figure}

Fig.~7 through Fig.~9 show the specific heat of the model for $S=1$, $3/2$ and $2$. The specific heat further 
accentuates the physics near the flattening of the entropy curves around $S_{max}/2$. At large anisotropy
there is clearly no such feature and the system only has a clear plateau at an entropy of $\frac{1}{2}\ln{2}$, where
the specific heat becomes vanishingly small. In contrast as one moves towards the isotropic limit, the specific heat
develops a three-peak structure. The highest temperature one corresponds to the flattening of the entropy curves
near $S_{max}/2$. The middle one corresponds to the entropy plateau at the value of $\frac{1}{2}\ln{2}$ and the lowest
one, not fully accessible to our numerical study, corresponds to the lifting of the degeneracy within the low energy
subspace. For $S=1$, the higher temperature peak extends down in $J_y$ values to $J_y=0.6$, where there is a clear flat region
in the specific heat. But, at this anisotropy, it goes away for higher spin, where the three-peak feature only arises
for $J_y=0.8$ or higher. For $S=2$, even at $J_y=0.8$ the higher temperature peak is becoming more of a flat top.
The data is again consistent with the idea that in the large-S limit, only the entropy plateau at $\frac{1}{2}\ln{2}$ will
remain as long as $J_y$ is not equal to unity. So, the isotropic limit is clearly singled out as being special.

\begin{figure}
\begin{center}
 \includegraphics[width=6cm]{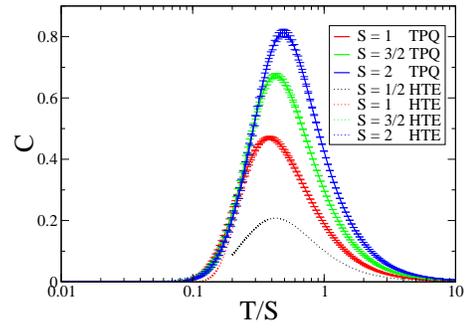}
\caption{\label{fig10} 
Specific heat of the anisotropic spin-S Kitaev model with $J_x=J_y=0.2$ and
$J_z=1$.
}
\end{center}
\end{figure}

The specific heat for different spin-values for $J_y=0.2$ are shown in Fig.~10. It is clear that the
entropy plateaus in this case correspond to the specific heat vanishing at intermediate temperatures.
The temperature scale for this goes as $JS$ as expected from the energy gap. Vanishingly small specific
heat is required for the plateaus to be sharply defined.

\section{Discussion of the numerical results and conclusions}

To understand the entropy plateaus we turn to the semiclassical limit. In the large-S limit,
any anisotropy $J_y<1$ causes the ground state to become collinear. Each spin must pair with 
its neighbor that couples the $z$ component of the spins and all spins must point along the
$z$ axis to obtain the lowest energy. There are $2$ ground states for each paired dimer of spins and the total number of ground
states is equal to $2^{N/2}$ for an N-site system. The gap to these states scales as $JS$.
This naturally explains the entropy plateaus
at $\frac{1}{2}\ln{2}$. The weak transverse couplings must ultimately lead to a further lifting of the degeneracy
in this subspace and for large S this must occur at very low temperature and would require much larger
system sizes to be valid in the thermodynamic limit,
which is beyond the reach of the present study.

For the isotropic spin-half model, the work by Nasu et al \cite{qmc1,qmc2} has shown that entropy plateaus arise just as nearest-neighbor spin correlations reach close to their ground state value. Since, nearest-neighbor spin correlations 
are proportional to the energy of the state, this is merely the statement that all the states contributing
to the entropy plateau have nearly the same energy. This result was also found to be true by Koga et al \cite{ktn} for higher
spin. It was also found by Nasu et al \cite{qmc1,qmc2} that the $Z_2$ flux variables averaged close to zero in the plateau region, implying that the flux variables were fully active in the plateau region. Only when the system transitions out of the plateau region and starts heading towards the zero entropy state at lower temperatures the $Z_2$ flux variables head to their ground state value of $+1$. This had the nice interpretation that the residual entropy of $\ln{2}/2$ corresponds to the number of flux configurations which is $2$ per hexagon, that is $2^{N/2}$ in total, where $N$ is number of sites.
Koga et al \cite{ktn} found that the flux variables average zero in the plateau region also for $S=1$ and only head towards their ground state value of unity when the system heads out of the plateau towards zero entropy. However, just the flux variables only have $\ln{2}/2$ different values, independent of spin. So, this cannot explain the larger value of entropy at the plateau for higher-S.

To understand the incipient entropy plateaus in the isotropic model with $S>1/2$ better we need to look for a much
larger low energy manifold in the classical limit. Indeed as shown by Baskaran et al. \cite{bss} and Chandra et al. \cite{crd}
the degeneracy in the classical limit for the isotropic model is significantly larger. Any dimer
covering of the lattice defines $2^{N/2}$ Cartesian ground states. Since the number of dimer coverings of
the honeycomb lattice has \cite{dimer} the asymptotic form $(1.381)^{N/2}$, this implies at least $(1.662)^N$
ground states of the system. But, these will lead to entropy less than $S_{max}/2$ even for $S=1$. Note that
$S_{max}$ diverges logarithmically as $S$ goes to infinity.

Baskaran et al.\cite{bss} also showed that the manifold of classical ground states has a continuous degeneracy.
To show this, let us decompose the lattice into disjoint parts by constructing a set of self avoiding walks and closed loops (self-avoiding polygons), such that every site belongs to one and only one walk or loop. 
In order to be able to dimerize each walk in two ways, there should not be any open ends in the bulk of the lattice.
%To see the larger degeneracy one can first consider drawing a curve through the lattice, where every site is
%connected to precisely two other sites. One can regard such a curve as a non-intersecting 
%path which decomposes into many disjoint closed self avoiding walks. 
To get the largest degeneracy one wants
the decomposition to have maximum number of disjoint pieces, and that is achieved by choosing loops around hexagons
as shown in Fig.~11. This 
represents a Plaquette Valence Bond State of the honeycomb lattice (See Fig.~11).

Each hexagonal plaquette can be dimerized in two ways and each dimerization defines $2^3$ Cartesian states,
giving a total of 16 Cartesian states for each hexagon.
If we considered only the Cartesian states these would imply a ground state degeneracy of ${16}^{N/6}$ and
a residual entropy of only approximately $0.462$. However as shown by Baskaran et al, there is a continuous
one-parameter family of ground states, which means the total number of ground states is not countable. 
To see this independent continuous degeneracy for each plaquette, consider the spin configurations characterized by
a parameter $\theta$ in the plaquette shown in Fig.~1b (assuming a ferromagnetic Kitaev model):

\begin{eqnarray}
\vec S_1 = \sin{\theta} \hat z + \cos{\theta} \hat y \nonumber\\
\vec S_2 = \sin{\theta} \hat z + \cos{\theta} \hat x \nonumber\\
\vec S_3 = \cos{\theta} \hat x + \sin{\theta} \hat y \nonumber\\
\vec S_4 = \sin{\theta} \hat y + \cos{\theta} \hat z \nonumber\\
\vec S_5 = \cos{\theta} \hat z + \sin{\theta} \hat x \nonumber\\
\vec S_6 = \sin{\theta} \hat x + \cos{\theta} \hat y \nonumber\\
\end{eqnarray}
One can easily verify that with arbitrary $\theta$ selected independently in each hexagon leaves the infinite system
in the classical ground state.
In the strictly classical limit, there are $(8\ \infty)^{N/6}$ ground states and hence an unbounded entropy per spin.
For finite $S$ this continuous degeneracy could give rise to a large degeneracy, possibly scaling with $S$
to some power, with an associated entropy that goes as $S_{max}/2$.

To look for hints of this numerically, we have studied the exact spectrum of a single hexagon plaquette for $S$ =$1$, $3/2$, $2$
and $5/2$. The idea is to look for a gap in the spectrum that persists in the large-S limit, and separates a lower
energy manifold of states from the rest. In that case, the $1/S$ corrections may still leave a meaningful low energy
manifold of states that corresponds to the incipient entropy plateau. 
%This situation is analogous to a half-filled Hubbard
%model, where $U/t$ is of order unity.

\begin{figure}
\begin{center}
 \includegraphics[width=7cm]{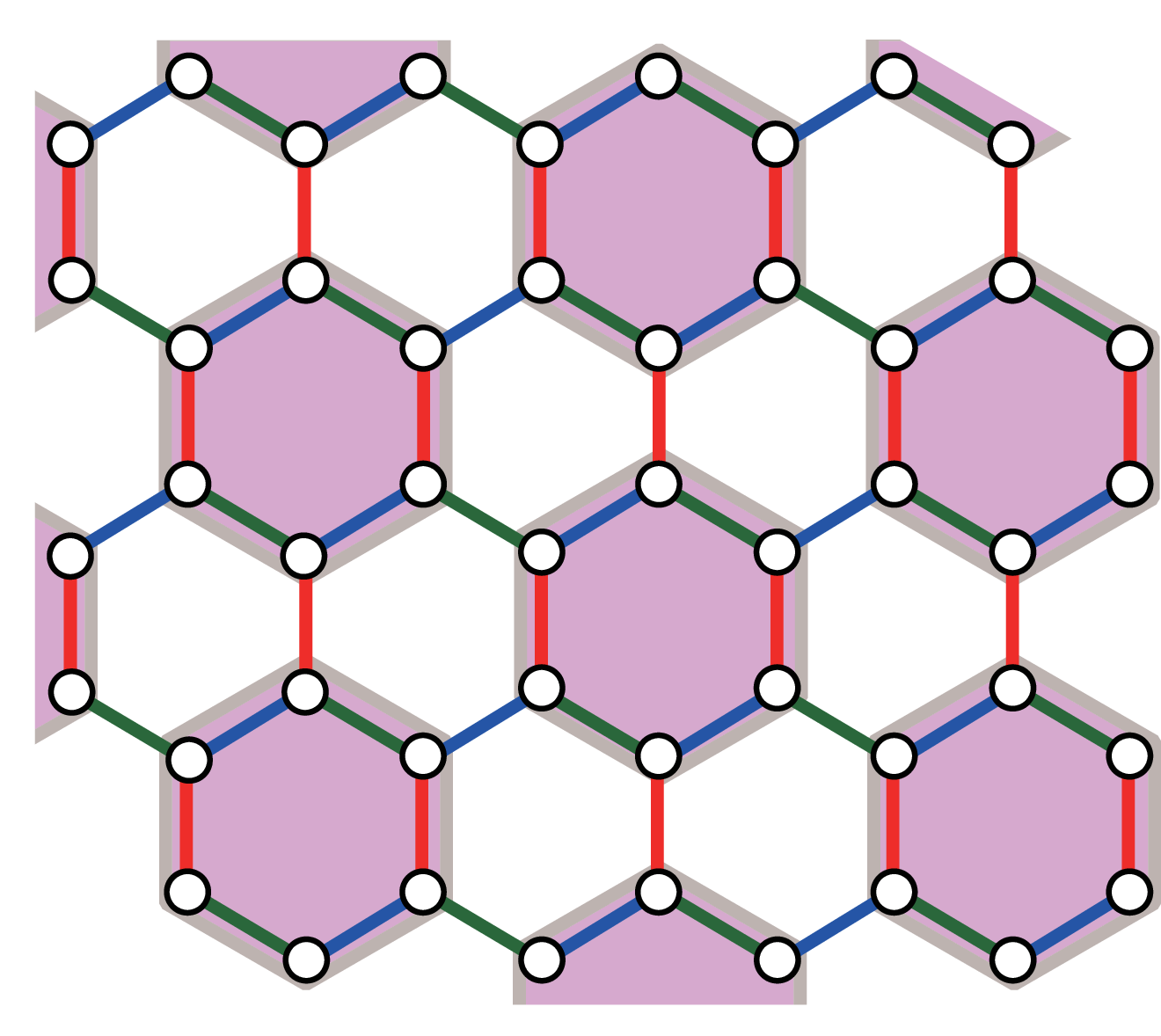}
\caption{\label{fig11} 
A set of closed loops on the honeycomb lattice shown by the 
thick grey lines that is also equivalent to a Plaquette Valence-Bond state.
}
\end{center}
\end{figure}

The Hilbert space dimension of a single hexagonal plaquette grows as $D=(2S+1)^6$.
In all cases, we find that there are many gaps in the spectrum, much larger than the typical energy
spacing. Large gaps are not unusual near
the ends of the spectrum. But, we find that the gaps in the spectrum also exist away from the ends of the spectrum. 
As shown in Table~1, the last prominent gap (furthest away from the ends of the spectra)
occurs approximately $D^{1/2}$ away from the edge. In other words
this gap defines a low energy manifold with an entropy which goes as $S_{max}/2$. If such gaps persist all the
way to the large-$S$ limit, they could imply a low energy manifold for the infinite system, with
an entropy equal to $S_{max}/2$.
We should note that gaps in the finite clusters need not imply a true gap in the thermodynamic limit,
but only a pseudogap or reduced density of states in the thermodynmic limit.
These gaps are analogous to those in half-filled Hubbard model at moderate $U/t$ showing the freezing of
charge degrees of freedom at an entropy of $\ln{2}$ per site.

\begin{table}[h]
\caption{
Gaps in the spectra of spin-S Kitaev Models. $D$ is the total Hilbert space dimension.
$\Delta_a=(E_{max}-E_{min})/D$ is the typical energy-level spacing, $D_l$ is the dimension
of the Hilbert space below the noted gap, $\Delta_l$ is the gap that separates $D_l$ states
from the rest of the system. The ratio $S_l/S_{tot}$ is the ratio of the entropy for
the low energy manifold to the total entropy.
}
\begin{tabular}{rrrrrr}
\hline\hline
S           & $D$          & $\Delta_a$ & $D_l$ & $\Delta_l$ & $S_l/S_{tot}$ \\
1           & 729          &  0.010     & 39    &  0.20      &  0.556 \\
3/2         & 4096         &  0.0037    & 64    &  0.36      &  0.5 \\
2           & 15625        &  0.0017    & 122   &  0.23      & 0.497 \\
5/2         & 46656        &  0.00086   & 232   &  0.27      & 0.507 \\
\hline\hline
\end{tabular}
\end{table}

We note that this analysis says nothing about further selection within this manifold, which could
proceed in the isotropic limit as discussed by Rousochatzakis et al \cite{rsp}. Also, there are three
ways to form Plaquette Valence Bond state on the lattice. $1/S$ corrections could restore the lattice symmetry by
mixing the very large number of degenerate states.

There is another
intriguing result of Baskaran et al. \cite{bss} that may be relevant to the incipient entropy plateau. They have shown
that there is a representation of the spins in terms of Majorana fermion operators for half-integer spins
that splits the $(2S+1)^N$ states into a $(S+1/2)^N$ states in direct product with $2^N$ states. Baskaran et al. \cite{bss}
argue that a modified hamiltonian may lead to a soluble model with (S+1/2) copies of Majorana Fermions. Clearly
a large number of copies of the Majorana Fermions can also lead to large entropy.
It would be very interesting if the incipient plateau reflects the onset of emergent Majorana variables.
Exploring such a connection is beyond the scope of this work.
It is interesting, however, that our numerical results suggest an incipient entropy plateau for both integer and half-integer
spins, where as the Majorana representation is realized in Baskaran et al work only for half-integer spins.
They are replaced by hard-core Bosons for integer spins. 
Both, in the case of fermions with point fermi-surfaces and bosons with
linear dispersion, one would obtain a $T^2$ entropy. The results in Fig.~2 are consistent with such a $T^2$ initial
correction above the entropy plateau.
Study of real time dynamics at and below the temperature
for the entropy plateau may throw further light on this emergenet subspace and the difference between possible fermionic
and bosonic excitations in half-integer and integer spins respectively.

It would be interesting to look for real materials that have dominant Kitaev exchange with $S> 1/2$. Given that
there are many effective ab initio approaches to designing spin-half Kitaev materials 
\cite{khaliullin,hykee,valenti,valenti2,motome,ybk,trebst},
it would not be surprising if higher spin Kitaev materials may also be discovered soon.

We would like to thank J. Nasu for valuable discussions and for sending us their Monte Carlo data for $S=1/2$ Kitaev system. Part of the numerical calculations were performed in the supercomputing systems in ISSP, the University of Tokyo.
The series expansion work was supported by the Australian Partnership for Advanced Computing (APAC) National Facility.
This work was supported by Grant-in-Aid for Scientific Research from JSPS, KAKENHI Grant Nos. JP18K04678, JP17K05536 (A.K.) and by the US National Science Foundation grant number DMR-1306048 (R. R. P. S.).

\end{document}